\documentclass[12pt,preprint]{aastex}

\DeclareGraphicsExtensions{.eps}

\begin{document}

\title{Signatures of hadronic cosmic rays in starbursts?\\
High-energy photons and neutrinos from NGC253}

\author{Gustavo E. Romero\footnote{Instituto Argentino de Radioastronom\'{\i}a
(IAR), C.C.\ 5, 1894 Villa Elisa, Argentina. E-mail:
romero@irma.iar.unlp.edu.ar} \& Diego F. Torres\footnote{Lawrence
Livermore National Laboratory, 7000 East Ave., L-413, Livermore,
CA 94550, USA. E-mail: dtorres@igpp.ucllnl.org}}

\begin{abstract}
We show that it appears possible for starburst galaxies, like the
nearby NGC 253, recently identified as a TeV
source by the CANGAROO collaboration, to emit a significant amount
of high-energy $\gamma$-rays and neutrinos through hadronic
processes in their cores. We suggest that proton illumination of
the inner winds of massive stars can be a viable mechanism for
producing TeV $\gamma$-rays and neutrinos without a strong MeV-GeV
counterpart. The rich stellar content of the starbursts, with
millions of early-type stars concentrated in the central regions,
where collective effects of the stellar winds and supernovae can
produce a significant enhancement of the cosmic ray density,
provides an adequate scenario for TeV $\gamma$-ray generation.
Close starbursts are also found to be potential sources for
km-scale neutrino telescopes, like ICECUBE, within reasonable
integration times.
\end{abstract}


\section{Introduction}

Very recently, the CANGAROO collaboration reported the
detection of the nearby starburst galaxy NGC 253 at TeV
$\gamma$-ray energies (Itoh et al. 2002). The purpose of this Letter is
to show that it is possible for starburst galaxies, like
NGC 253, to produce TeV photons and neutrinos through hadronic processes.

\section{A plausible scenario}

Collective effects of strong stellar winds and supernova
explosions in star-forming regions are expected to result in
particle acceleration up to multi-TeV energies, or even beyond
(e.g. Montmerle 1979; Cass\'{e} \& Paul 1980; Bykov \& Fleishman
1992a,b; Anchordoqui et al. 1999; Bykov 2001). Recently, the
region co-spatial with the Cygnus OB2 association has been
detected as a TeV $\gamma$-ray source by the HEGRA telescope array
(Aharonian et al. 2002). TeV cosmic rays (CRs) accelerated in the
association might be responsible for the high-energy $\gamma$-ray
emission through the hadronic illumination of some suitable target
(Butt et al. 2003). A nearby EGRET source (3EG J2033+4118)
has also a likely stellar origin (Chen et al. 1996, Romero et al.
1999, Benaglia et al. 2001).

In a starburst, where the ambient density of CRs is enhanced by a
local high rate of supernova explosions and the collective effects
of strong stellar winds (e.g. Bykov 2001), $\gamma$-rays can be
produced by interactions of relativistic protons with the rich
interstellar medium. This interaction produces neutral pions that
quickly decay into $\gamma$-rays. The latter can also be produced
by leptonic processes, like inverse Compton interactions with the
strong FIR field, and relativistic bremsstrahlung in the ambient
gas. All these processes have been modelled with a set of
reasonable parameters for NGC 253 by Paglione et al. (1996). The
expected total $\gamma$-ray flux, however, is below the EGRET
sensitivity limit, in accordance with the non-detection of the
galaxy in the MeV-GeV range (Bhattacharya et al. 1994, Blom et al.
1999).

A source of $\gamma$-rays not previously considered in the
mentioned analysis is the production at the base of the strong
stellar winds of early-type stars. When the relativistic particles
are accelerated in the stellar wind itself, for instance through
multiple shocks produced by line-driven instabilities (e.g. White
\& Chen 1992), the $\gamma$-ray flux from pion decays is limited
to MeV energies. However, if the star is immersed in an external
bath of relativistic particles accelerated at larger scales, e.g.
in the core of the starburst, TeV protons can penetrate the base
of the wind to produce TeV $\gamma$-rays. The injection of
MeV---GeV particles into the wind of the stars will be strongly
attenuated by the modulation effects of the wind itself.

The dense medium in which these $\gamma$-rays are produced is,
nonetheless, transparent to $\gamma$-ray propagation. The optical
depth to pair production is $\tau \sim n\,\sigma\,R$, where $n$ is
the photon number density, $\sigma$ is the photon-photon cross
section, and $R$ is the distance that the photon must travel to
escape. In order to compute $n$, we need the photon energy density
at the base of the stellar wind, $U$. We shall assume that target photons in the
wind have typical energies of $\sim$ 1~eV. Then,
$U=3\,L/(4\pi\,R^2\,c)\sim 2\times 10^{10}$~eV~cm$^{-3}$, where we
have taken, for order-of-magnitude estimates, the following
typical values: $R\sim 5\times 10^{14}$~cm and $L\sim
10^{39}$~erg~s$^{-1}$. Then, $n=U/(1~\rm eV)$, and the optical depth
is very small, $\tau\sim 2 \times 10^{-5}$. The computation
of $\tau$ makes use of the $\gamma-\gamma$ (1 TeV---1eV) cross
section, $\sigma \sim 2 \times 10^{-30}$ cm$^2$ (Lang 1999). Once
TeV photons are produced and escape from the wind region, they must yet leave
the galaxy to reach the Earth. The starburst has a FIR field with $U
\sim 200$ eV cm$^{-3}$ (e.g. Paglione et al. 1996), and  photons
have to travel distances of about 70-100 pc to leave the active
region of the galaxy. The optical depth is, in this case, $\tau =
(200$ photons cm$^{-3})\; \times (2\times 10^{-30}$ cm$^{2})\;
\times (100$ pc) $\approx$ 1.2 $\times 10^{-7}$, which is also small
enough to secure that photons can escape from the galaxy.


\section{NGC 253}

NGC 253 has been described as an archetypal starburst galaxy by
Rieke et al. (1980), and it has been extensively studied from
radio to $\gamma$-rays (e.g. Beck et al. 1994, Paglione et al.
1996, Ptak 1997).
The supernova rate is estimated to be as high
as $0.2-0.3$ yr$^{-1}$, comparable to the massive star formation
rate, $\sim 0.1$M$_\odot$ yr$^{-1}$ (Ulvestad et al. 1999, Forbes
et al. 1993). The central region of this starburst is packed with
massive stars. Watson et al. (1996) have discovered four young
globular clusters near the center of NGC 253; they alone can
account for a mass well in excess of 1.5$\times 10^6 M_\odot$ (see
also Keto et al. 1999). Assuming that the star formation rate has
been continuous in the central region for the last 10$^9$ yrs, and
a Salpeter IMF for 0.08-100 $M_\odot$, Watson et al. (1996) find
that the bolometric luminosity of NGC 253 is consistent with 1.5
$\times 10^8 M_\odot$ of young stars. Based on this evidence,
it appears likely that there are at least tens of millions
of young stars in the central region of the starburst.
Many stars
might be obscured by the large amount of molecular material, as in
the case of the massive Galactic cluster Westerlund 1 (Clark and
Negueruela 2002).

Physical, morphological, and kinematic evidence for the existence
of a galactic superwind has been found for NGC 253 (e.g. McCarthy
et al. 1987, Heckman et al. 1990). This superwind creates a cavity
of hot ($\sim10^8$ K) gas, with cooling times longer than the
typical expansion time scales. As the cavity expands, a strong
shock front is formed on the contact surface with the cool
interstellar medium.
The shock velocity can reach thousands of kilometers per second,
and iron nuclei can be efficiently accelerated up to high energies
$\sim 10^{20}$ eV by diffusive mechanisms operating at large
scales. Indeed, NGC 253 has been proposed as the origin for some
of the observed ultra-high energy CRs (Anchordoqui et al. 1999,
2002).


The CANGAROO collaboration did not give a precise photon spectrum
for NGC 253. The reported data were only the integrated flux above
0.5 TeV (securing an 11$\sigma$ confidence detection) and the
possible range of power law indices $\Gamma=$(1.85, 3.75) (Itoh et al.
2002). We can summarize these results expressing the differential
photon spectrum as:
$
 F_{\gamma}(E_{\gamma})  =
  B\,\, \left({ E_\gamma}/{1\, {\rm TeV}}\right)^{-\Gamma} \,\,\,\,
  {\rm cm^{-2}\, s^{-1}\, TeV^{-1}},
 $
where  $B   =  (3.8\pm 1.2) \times 10^{-12} $ for $\Gamma=1.8$, or
$B = (3.2\pm 1.0) \times 10^{-12} $ for $\Gamma=3.75$. It is worth
noticing that the direct extrapolation of the spectrum at TeV
energies down to the GeV band would yield an integrated value
compatible with the EGRET upper limits for NGC 253 only for power
law indices $\Gamma \lesssim 2$. Spectra softer than those with
$\Gamma=2$ at TeV energies should present a break at a few
hundred GeV.

\section{Hadronic origin of the TeV emission from NGC 253?}

The differential $\gamma$-ray photon number distribution from
neutral pion decays at the source is given by (e.g. Gaisser 1990)
$ N_{\gamma}(E_{\gamma})= N_p(E_\gamma)\, \xi_A \, 4 \pi\,
\sigma_{pp}(E_\gamma) \, N \, {2Z_{p\rightarrow
\pi^0}^{(\alpha)}}/{\alpha}. $ Here, $E_\gamma$ is the photon
energy, $N=\Sigma/m_p$ is the column density, $m_p$ is the proton
mass, and $\alpha$ is the proton spectral index, such that
$N_p=K_p E^{-\alpha}$. In addition, $\xi_A\sim 1.5$ is a
correction factor that takes into account possible effects of
heavier nuclei, $\sigma_{pp}(E_\gamma)\sim 35$ mb is the hadronic
interaction cross section, and $Z_{p\rightarrow \pi^0}^{(\alpha)}
\sim 0.17$ stands for the fraction of kinetic energy
of the parent proton that is transferred to the neutral pion
during the collision. For a photon/proton spectral index $ \alpha
\sim 2$ we get $N_\gamma(E_{\gamma})=1.12\times 10^{-25} K_p
E_{\gamma}^{-2}$ cm$^{-3}$ TeV$^{-1}$ (notice that 1 erg $\sim$ 1
TeV). Assuming typical grammages for the stellar winds of OB stars
in the range $\Sigma \sim 50-150$ g cm$^{-2}$ (e.g. White 1985),
we get $N_\gamma(E_{\gamma})=(3.4-10.1) K_p E_\gamma^{-2}$
cm$^{-3}$ TeV$^{-1}$ . The luminosity in the CANGAROO observing
range is
\begin{eqnarray}
L &=& 4 \pi R^2 c \int_{\sim 0.5{\rm TeV}}^{\sim 20{\rm TeV}}
N_\gamma(E_\gamma) E_\gamma dE_\gamma \\ \nonumber &=& (4.7-13.9)
\times 10^{12} R^2 K_p\; {\rm erg}\; {\rm s}^{-1},
\end{eqnarray}
where $R$ is the size of the base of the wind and $c$ is the
speed of light (all numerical coefficients are for cgs units). The
CR energy density is, on the other hand,
\begin{equation}
\omega_{\rm CR} =  \int_{\sim 1{\rm GeV}}^{\sim 20{\rm TeV}}
N_p(E_p) E_p dE_p = 9.9 K_p\; {\rm erg}\; {\rm s}^{-1} \equiv
\varsigma \omega_{\rm CR,\odot},
\end{equation}
where $\varsigma$ is the amplification factor of the CR energy
density with respect to the local value, $ \omega_{\rm CR,\odot}\sim
10^{-12}$ erg cm$^{-3}$. This implies that the TeV luminosity can
be written as $L=(0.47-1.4) R^2 \varsigma$ erg s$^{-1}$. The
latter corresponds to the luminosity of a single star illuminated
by TeV CR protons. In the central region of the starburst,
millions of stars could be illuminated at the same time. The total
luminosity would then be $ L_{\rm
total}=\sum_{i=1}^{N_\star} (0.47-1.4)\, R_i^2 \, \varsigma_i \;
{\rm erg}\; {\rm s}^{-1},
$ where $N_\star$ is the number of contributing
stars, and we have allowed for the possibility that not all of them will
have the same $R$, nor will they be located within regions having
the same CR enhancement factor.

\begin{figure*}[t]
\centering
\includegraphics[width=0.3\textwidth,height=7cm]{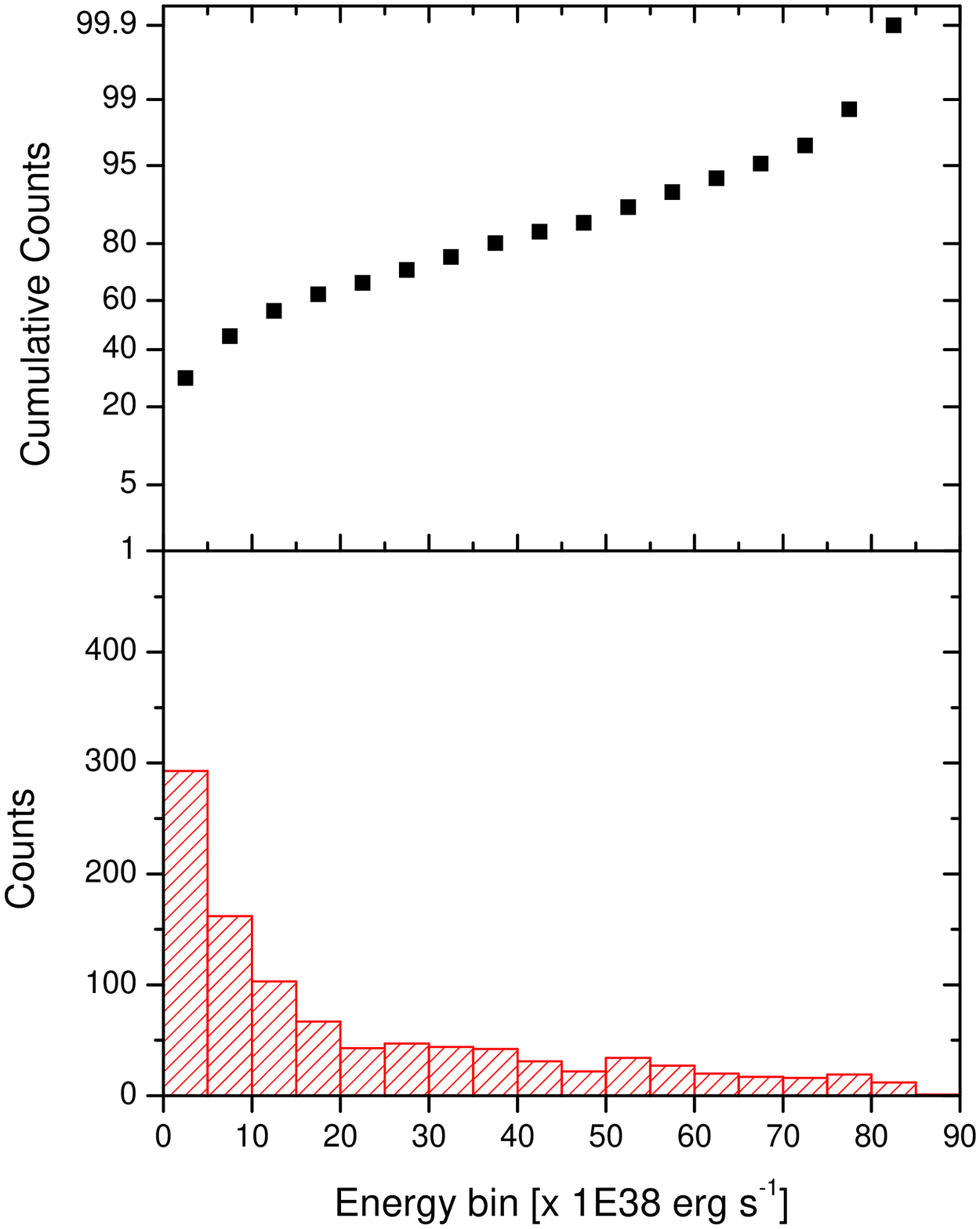}\hspace{2cm}
\includegraphics[width=0.3\textwidth,height=7cm]{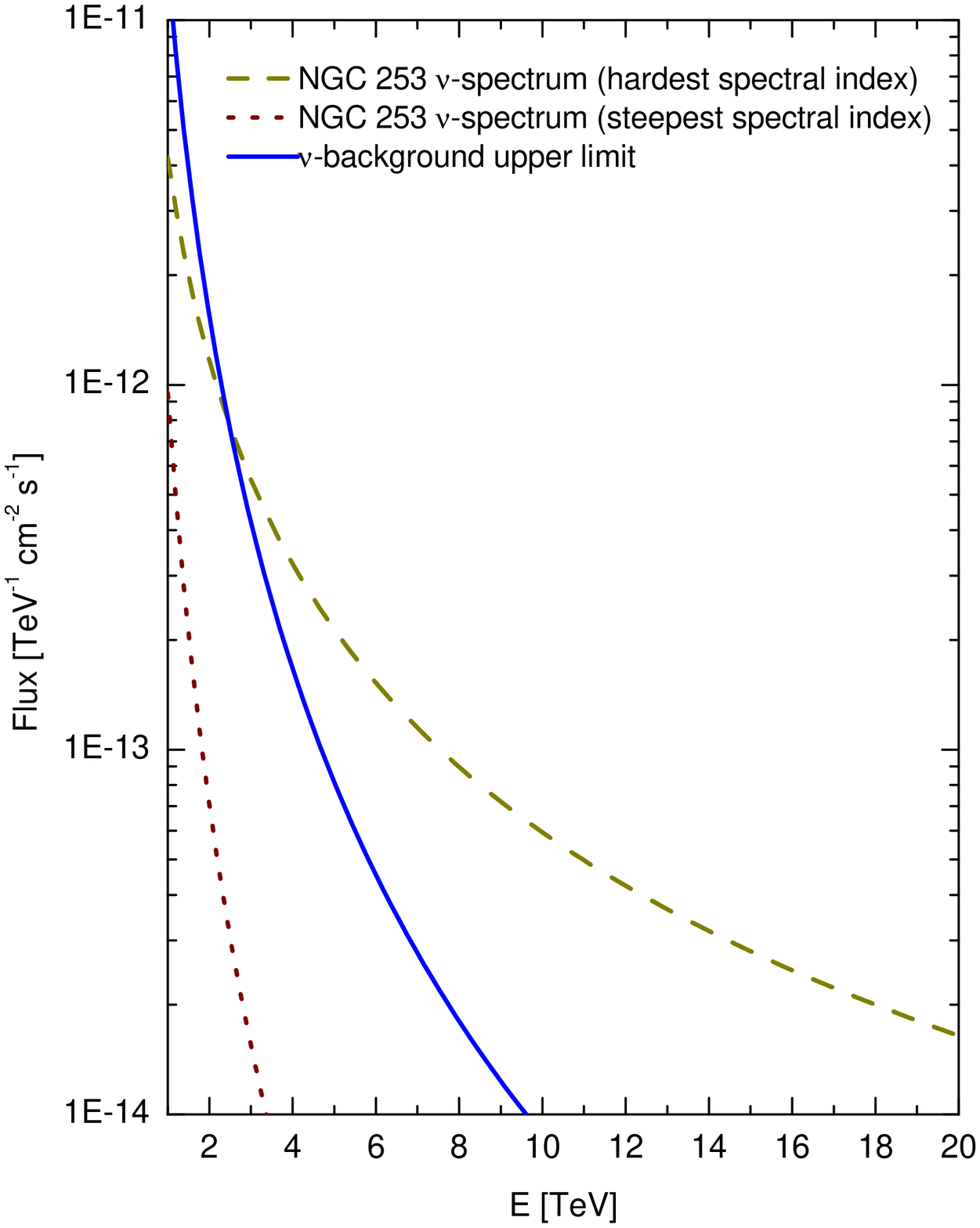}
\caption{Left (a): Results of 1000 Monte Carlo trials to compute
the total TeV luminosity of a starburst lying at the distance of
NGC 253. Right (b): The hadronic
model expectation of neutrinos from NGC 253 compared with the
upper limit for the neutrino background in a ${1^\circ \times
1^\circ}$-generation neutrino telescope.} \label{fig2}
\end{figure*}

If we now assume the following ranges for the different parameters
involved: $1\times 10^{12}<R[{\rm cm}]<1\times 10^{15}$ (Lamers
1999), $300 < \varsigma < 3000$ (Suchkov et al. 1993), and
$10^6<N_\star<5 \times 10^7$ (see above), there is ample room for
the total luminosity to reach values above 10$^{39}$ erg s$^{-1}$,
i.e. the observed CANGAROO luminosity above 0.5 TeV, given the
distance to NGC 253 (2.5 Mpc, de Vacoulers 1978). To show this, we
have carried out Monte Carlo numerical simulations where the
number of stars illuminated by the accelerated protons, the
grammage, the size of the base of the wind for each star, and
the ambient enhancement of CRs are independently allowed to take values
within their assumed ranges. For each of these cases, the total
luminosity was computed. We show the results in Fig. 1a, for
a sample of 1000 trials. Only for less than 40\% of
the trials, the total luminosity is below 10$^{39}$ erg s$^{-1}$.
This scenario then appears to be a plausible explanation for the
TeV observations of NGC 253.

Note that multi-TeV electrons in the starburst central region will
experience inverse Compton losses with the FIR fields (typical
energy density $\sim 200$ eV cm$ ^{-3}$) in the Klein-Nishina
regime, which result in a very steep $\gamma$-ray spectrum (that
can be approximated by power-law indices $\gtrsim 3.2$ (although
notice that the actual spectrum is not a pure power law, see e.g.
Georganopoulos et al. 2001). The total leptonic contribution from
the central starburst region at TeV energies is then negligible in
comparison to the hadronic one.

MeV---GeV particles cannot easily penetrate the
stellar winds, and hence these winds will glow mainly at TeV
energies. The modulation of CRs by stellar winds is a complex
subject that has been studied in detail only in the context of the
relatively weak solar wind.
In such case,
three-dimensional models including diffusion and the effects of
the terminal shocks have been developed (e.g. K\'ota \& Jokipii
1983, Jokipii et al. 1993). These models can explain fairly well
the basic observed features of the nucleonic CR component in the
solar system, where strong gradients in the CR density are known
to exist from interplanetary spacecraft measurements (e.g. Quenby
et al. 1990).\footnote{Although it is highly dependent of the solar
activity and the inclination angle of the incident particles, the
proton flux decreases to less than $\sim 30$\% of its initial
value at 80 A.U. for 2.5 GeV particles, and to less than $\sim
10$\% for 250 MeV particles.}
In OB stars, with extremely
supersonic winds and mass loss rates orders of magnitude higher
than the solar values, these effects should be far more
pronounced, leading to an almost complete rejection of externally
injected MeV---GeV particles at the base of the winds and thus to
an absence of a strong (above EGRET sensibility) $\gamma$-ray
counterpart at these energies. The modulation effect depends on
the parameter $\epsilon \sim u R_{\rm w}/D$,
  where $u$ is the
  stellar wind velocity, $R_{\rm w}$ is the radius of the wind shell, and $D$ is the diffusion
  coefficient. $\epsilon$
  measures the ratio between the diffusive timescale of the particle to their convective timescale.
  The former should be shorter than the latter for a TeV particle to enter into the wind cavity.
  For a typical O star,
  $R_{\rm w}> 10$ pc and $u\sim 2500$ km/s. Hence,
  the diffusion coefficient of the TeV particles should be at least $10^{28}$ cm$^{2}$ s$^{-1}$,
  similar to what is estimated for our Galaxy (e.g. Berezinskii et al. 1990).

\section{Neutrino signal}

If the $\gamma$-ray emission is explained by the decay of neutral
pions, their charged counterparts must produce a neutrino flux.
Following Alvarez-Mu\~niz \& Halzen (2002), the $\nu_\mu +
\bar\nu_\mu$ neutrino flux $F_\nu(E_\nu)$ produced by the decay of
charged pions in the source can be derived from the observed
$\gamma$-ray flux $F{_\gamma}(E_\gamma)$ by imposing energy
conservation (see also Stecker 1979):
\begin{equation}
\int_{E_{\gamma}^{\rm min}}^{E_{\gamma}^{\rm max}} E_\gamma
F{_\gamma}(E_\gamma) dE_\gamma = K \int_{E_{\nu}^{\rm
min}}^{E_{\nu}^{\rm max}} E_\nu F_\nu(E_\nu) dE_\nu~.
\end{equation}
Here, ${E_{\gamma\; [\nu]}^{\rm min}}$ ($E_{\gamma \; [\nu]}^{\rm
max}$) is the minimum (maximum) energy of the photons (neutrinos)
and the pre-factor $K=1$.\footnote{Note that there are two
muon-neutrinos (out of the pion-to-muon-to-electron decay chain)
for every photon produced. Each of the latter carries, on average,
an energy $E_\pi/4$, whereas each photon carries on average half
the energy of the pion. Therefore, there is as much energy in
photons as there is in neutrinos, and $K=1$.}
The maximum neutrino energy is fixed by the maximum energy of the
accelerated protons ($E_p^{\rm max}$) which can in turn be
obtained from the maximum observed $\gamma$-ray energy
($E_{\gamma}^{\rm max}$) as: $ E_p^{\rm max}\sim 6
E_{\gamma}^{\rm max}\;,\;\;E_{\nu}^{\rm max}\sim {1 / 12} E_p^{\rm
max} $ (Alvarez-Mu\~niz \& Halzen 2002).
The minimum neutrino energy is fixed by the threshold for pion
production,
$
E_p^{\rm min}=\delta ~ {(2m_p+m_\pi)^2-2m_p^2 \over 2m_p},
\label{ethpp}
$
where $\delta$ is the Lorentz factor of the accelerator relative
to the observer.
Assuming a Lorentz factor of order 1, the $\nu$-spectrum results
in $F_{\nu}(E_\nu)\sim 4.2 \times 10^{-12} (E/{\rm TeV})^{-1.85}$
TeV$^{-1}$ cm$^{-2}$ s$^{-1}$ for the harder photon index, and
$F_{\nu}(E_\nu)\sim 9.5 \times 10^{-13} (E/{\rm TeV})^{-3.75}$
TeV$^{-1}$ cm$^{-2}$ s$^{-1}$ for the softer photon index.

The event rate of atmospheric
$\nu$-background is (see Anchordoqui et al. 2002b)
\begin{equation}
\left. \frac{dN}{dt}\right|_{\rm B} = A_{\rm eff}\, \int dE_\nu
\,\frac{d\Phi_{\rm B}}{dE_\nu}\, P_{\nu \to \mu}(E_\nu)\,\,\Delta
\Omega_{1^\circ \times 1^\circ}\,, \label{background}
\end{equation}
where $A_{\rm eff}$ is the effective area of the detector, $\Delta
\Omega_{1^\circ \times 1^\circ} \approx 3 \times 10^{-4}$~sr is an
assumed bin size for the observation (ICECUBE-generation detector,
Karle et al. 2002), and $d\Phi_{\rm B}/dE_\nu \lesssim 0.2
\,(E_\nu/{\rm GeV})^{-3.21}$~GeV$^{-1}$ cm$^{-2}$ s$^{-1}$
sr$^{-1}$  is the $\nu_\mu + \bar \nu_\mu$ atmospheric
$\nu$-flux~(Volkova 1980, Lipari 1993). Here, $P_{\nu \to \mu}
(E_\nu) \approx 1.3 \times 10^{-6} \,(E_\nu/{\rm TeV)}^{0.8}$
denotes the probability that a $\nu$ of energy $E_\nu
> 1~{\rm TeV}$, on a trajectory through the detector, produces a
muon~(Gaisser et al. 1995). On the other hand, the $\nu$-signal is
\begin{equation}
\left. \frac{dN}{dt}\right|_{\rm S} =  A_{\rm eff} \,\int dE_\nu
\, F_{\nu}(E_\nu)\,P_{\nu \to \mu}(E_\nu)\  \,.
\label{yellowsubmarine}
\end{equation}
The signal-to-noise (S/N) ratio in the 1--20 TeV band is then $(
\left. {dN}/{dt}\right|_{\rm S})/(\sqrt{\left.
{dN}/{dt}\right|_{\rm B}})=0.1-2.5$ for one year of operation,
depending on the photon spectral index. A NEMO-generation
detector, with a bin size $\Delta \Omega\sim {0.3^\circ \times
0.3^\circ}$ (Riccobene et al. 2002), would have a S/N ratio
roughly three times larger for the same integration time. The
effect of neutrino oscillations may reduce the flux by a factor of
three at Earth (Bilenky et al. 1999). However, for some possible
values of the photon spectral index, this signal could be detected
in a reasonable time span. The possible starburst neutrino spectra
producing the previously quoted S/N ratio are plotted, together
with an upper limit for the atmospheric neutrino background with
vertical incidence, in Fig.~1b.\footnote{Regrettably, the south
pole location of the ICECUBE telescope would render its
observation of NGC 253 nearly impossible; only neutrinos going
through Earth can be identified. The smaller northern telescope
ANTARES (0.1 km$^2$, Becherini et al. 2002) would need much longer
integration times to be able to distinguish NGC 253 from the
background, if at all possible. Nevertheless, the analysis
presented here can be directly applied to M82, a northern-sky
starburst at the same distance as NGC 253.}

\section{Concluding remarks}

Through hadronic interactions in their central regions, nearby
starburst galaxies can generate high-energy $\gamma$-ray emission
detectable at Earth. Proton illumination of the inner winds of
massive stars in a high density CR environment may result in TeV
sources without strong counterparts at MeV---GeV energies. The
high number of early-type stars with strong stellar winds in this
kind of galaxies might produce a strong collective effect,
providing the setting for accelerating protons up to multi-TeV
energies. In addition, we have found that starbursts might also be
potential sources for km-scale neutrino telescopes within
reasonable integration times, depending on the photon spectral
index. The detection ---or non-detection--- of the neutrino signal
may be an essential piece of evidence in the determination of the
origin of the high-energy radiation.

Gamma-rays from neutral pion decays will be produced not only in
the stellar winds at the core of the starburst, but also in the
diffuse matter among the stars. If we model the starburst region
as a disk with a radius of $\sim300$ pc and a thickness of
$\sim70$ pc, where the average density of H atoms in the diffuse
ISM is $\sim 300$ cm$^{-3}$ (Paglione et al. 1996), we get that
the expected flux above 1 TeV is $F(E>1\;{\rm TeV})\simeq 3.9
\times 10^{-13}$ ph cm$^{-2}$ s$^{-1}$, for an average CR
enhancement of $\varsigma\sim1000$ (see Torres et al. 2002 for
details of calculation). This means an additional diffuse
contribution of $\sim3\times10^{38}$ erg s$^{-1}$ to the total TeV
$\gamma$-ray luminosity of the starburst. In most of our models
this diffuse contribution is only a small fraction of the total
$\pi^0$-decay $\gamma$-ray emission, which is dominated by
interactions occurring in the stellar winds.

Finally, it is important to remind that the CANGAROO observations
are consistent with an extended source, whose size might be even
bigger than the galaxy itself. If this is true, it would imply
that there might be another component generating TeV radiation,
related with the galactic halo. This component might probably be
the result of Inverse Compton scattering of CMB photons by TeV
electrons accelerated at the superwind. In such case, since the
interactions occur in the Thomson regime, the spectrum should be
harder than that produced by hadrons in the center of the
starburst.\footnote{After the submission of this paper we were
informed that Itoh et al. (2003) discussed in detail this point.
We thank Prof. Yanagita for calling our attention to this.}

\acknowledgements

Comments by L. Anchordoqui, P. Benaglia, Y.M. Butt, C. Mauche,
M.M. Kaufman-Bernad\'o, M. Pohl, and an anonymous referee are
gratefully acknowledged. G.E.R.'s research is mainly supported by
Fundaci\'on Antorchas. Additional support was provided by CONICET
(under grant PIP N$^o$ 0430/98) and ANPCT (PICT 03-04881). The
work of D.F.T. was performed under the auspices of the U.S. DOE
(NNSA), by UC's LLNL under contract No. W-7405-Eng-48. His
research was done while he was a Visiting Scientist at SISSA
(Italy) and a Visiting Professor at the University of Barcelona
(Spain), to which he gratefully acknowledge their hospitality.


\end{document}